\documentclass[twocolumn,showpacs,preprintnumbers,amsmath,amssymb]{revtex4}
\usepackage{graphicx}% Include figure files
\usepackage{dcolumn}% Align table columns on decimal point
\usepackage{bm}% bold math

\newcommand{\bra}[1]{\langle#1\rvert}
\newcommand{\ket}[1]{\lvert#1\rangle}
\newcommand{\avr}[1]{\langle#1\rangle}
\renewcommand{\vec}[1]{\bm{\mathrm{#1}}}

\expandafter\ifx\csname
 natexlab\endcsname\relax\def\natexlab#1{#1}\fi
 \expandafter\ifx\csname bibnamefont\endcsname\relax
   \def\bibnamefont#1{#1}\fi
 \expandafter\ifx\csname bibfnamefont\endcsname\relax
   \def\bibfnamefont#1{#1}\fi
 \expandafter\ifx\csname citenamefont\endcsname\relax
   \def\citenamefont#1{#1}\fi
 \expandafter\ifx\csname url\endcsname\relax
   \def\url#1{\texttt{#1}}\fi
 \expandafter\ifx\csname
 urlprefix\endcsname\relax\def\urlprefix{URL }\fi
 \providecommand{\bibinfo}[2]{#2}
 \providecommand{\eprint}[2][]{\url{#2}}

\begin{document}

\title{Multimode cavity-assisted quantum storage via continuous phase-matching control}

\author{Alexey Kalachev${}^{1,2}$ and Olga Kocharovskaya${}^2$}
\affiliation{${}^1$Zavoisky Physical-Technical Institute of the Russian
Academy of Sciences, Sibirsky Trakt 10/7, Kazan, 420029, Russia,\\ ${}^2$Institute for Quantum Studies and Department of Physics, Texas A\&M University, College Station, TX 77843--4242,
USA}%

\date{\today}% It is always \today, today,

\begin{abstract}
A scheme for spatial multimode quantum memory is developed such that spatial-temporal structure of a weak signal pulse can be stored and recalled via cavity-assisted off-resonant Raman interaction with a strong angular-modulated control field in an extended $\Lambda$-type atomic ensemble. It is shown that effective multimode storage is possible when the Raman coherence spatial grating involves wave vectors with different longitudinal components relative to the paraxial signal field. The possibilities of implementing the scheme in the solid-state materials are discussed.
\end{abstract}


\maketitle

\section{Introduction}

Developing optical quantum memories is an important part of quantum optics
and quantum information \cite{LST_2009,HSP_2010,TACCKMS_2010,SAAB_2010}. In particular, storage and retrieval of single photons is expected to be necessary for creating scalable linear-optical quantum computers, realizing long-distance quantum key distribution via quantum repeaters and making deterministic single-photon sources. Memory devices which could store single-photon wave-packets with close to 100\%
efficiency and fidelity, and provide long and controllable storage
times and delay-bandwidth products, are demanded for practical quantum
information applications. An efficient storage has been demonstrated recently in gases \cite{HSCLB_2011} (87\% efficiency) and rare-earth-ion-doped solids \cite{HLLS_2010} (69\% efficiency) using the gradient echo memory (GEM) technique \cite{ALSM_2006}, which is a variant of controlled reversible inhomogeneous broadening protocol \cite{MK_2001,NK_2005}. Significant experimental progress has also been achieved in the framework of other approaches based on atomic frequency comb (AFC) \cite{ASRG_2009}, electromagnetically induced
transparency (EIT) \cite{FL_2000}, and off-resonant Raman interaction \cite{NWRSWWJ_2007}.
In particular, 43\% \cite{PGN_2008} and 78\% \cite{CLWDCCY_2013} were obtained in hot and cold atoms, respectively, using EIT, and 56\% was achieved using AFC in a rare-earth doped crystal placed in a cavity \cite{SLKR_2013}, a delay-bandwidth product of 2500 was demonstrated for storage of 300-ps pulses using off-resonant Raman scheme in a warm atomic vapor \cite{RMLNLW_2011}, and a bandwidth of 5~GHz was achieved via AFC in a doped waveguide \cite{SSJSOBGRST_2011}.

Storage and retrieval of an optical pulse is usually
accomplished by an appropriate amplitude modulation
of the control field or by using inhomogeneous
broadening of the resonant transitions. In the first case, control-field modulation should match the input pulse,
while in the second case we need to control atomic frequencies or to create artificial atomic structures.
In the present work, we develop another approach which requires neither inhomogeneous broadening nor temporal modulation of the control-field amplitude, but resorts to continuous phase-matching control in an extended resonant medium. We consider off-resonant Raman interaction of a single-photon wave packet and a classical control field in a three-level atomic medium. Under such conditions the phase-matching control can be achieved by modulating the refractive index of the resonant medium \cite{KK_2011,CHS_2012} or by modulating the direction of propagation of the control field \cite{ZKK_2013}. In any case, a continuous change of the wave vector of the control field during the interaction leads to the mapping of a single-photon state into a superposition of atomic collective excitations with different wave vectors (Raman spatial coherence grating) and vice versa.

In comparison with \cite{ZKK_2013}, where the free-space model of quantum memory was considered, here we discuss a cavity model. Enclosing an atomic ensemble in a cavity makes it possible to achieve high efficiency of quantum storage with optically thin materials. This may be especially useful for considered off-resonant Raman interactions, since the cross section of the two-photon transition is usually small. In addition, especially in the case of a transverse control field, reducing linear sizes of the atomic ensemble in a cavity allows one to significantly reduce the power of the control field. On the other hand, compared to \cite{KK_2011}, we develop a three-dimensional theory, which allows us to consider storage and retrieval of spatially multimode states containing information not only in the pulse envelope but also in the transverse profile of the field. The spatial multimode storage is crucially important for multiplexing in quantum repeaters \cite{CJKK_2007,SRASZG_2007}, which can significantly increase the rate of quantum communication in possession of short-time quantum storage, and for holographic quantum computers \cite{TNM_2008}. It was experimentally demonstrated using EIT \cite{VCH_2008,SFPRD_2008,HRBH_2010,DWZLSZG_2013} and GEM \cite{HSRPHLB_2012,GCMZL_2012}. The continuous phase-matching approach developed here is closely related to the quantum holographic storage \cite{VSP_2010} differing in that control-field angular scanning is used as the only resource for storage and retrieval of a spatial-temporal structure of weak optical pulses.

The paper is organized as follows. In Sec. II, we present the model and derive basic equations describing multimode cavity-assisted off-resonant Raman interaction. In Sec. III, storage and retrieval of multimode single-photon wave packets
is considered. In Sec. IV, we discuss the possibility of implementing the scheme in the solid-state materials. Section V concludes the paper with final remarks.

\section{The model and basic equations}

We consider a system of $N\gg 1$ identical three-level atoms which
are placed in a single-ended ring cavity and interact with a weak signal field (single-photon wave packet) to be stored and with a
strong control field (Fig.~1).
The atoms have a
$\Lambda$-type level structure, and the fields are Raman resonant to
the lowest (spin) transition. We restrict the consideration of the cavity field to a single longitudinal mode. On the other hand, the cavity volume is supposed to have a large Fresnel number, which allows us to consider different transverse modes. In what follows, we assume the cavity to be formed by rectangular mirrors with cross section $A=L_xL_y$ and use a set of the mode functions
\begin{equation}\label{U}
u_{mnp}(\vec{r})=e^{iq_pz} u_{mn}(\vec{r}),
\end{equation}
where $q_p=2\pi p/L$, $p\in \mathbb{Z}$, satisfying the conditions of completeness and orthogonality
\begin{gather}
\frac{1}{V}\sum_{mnp}u_{mnp}^\ast(\vec{r})\,u_{mnp}(\vec{r}')=\delta(\vec{r}-\vec{r}'),\\
\frac{1}{V}\int d\vec{r}\, u_{mnp}^\ast(\vec{r})\,u_{m'n'p'}(\vec{r})=\delta_{mm'}\,\delta_{nn'}\,\delta_{pp'}.
\end{gather}
Here $V=AL$ is the cavity volume, and $L$ is the total cavity length. The coordinate system originates at the center of the atomic system. The paraxial signal field in the cavity corresponding to a single longitudinal mode is written as
\begin{equation}
E_s(\vec{r},t)=i\sqrt{\frac{\hbar\omega_s}{2\varepsilon_0V}}\sum_{mn}\,u_{mn}(\vec{r})\,a_{mn}(t)\,e^{ik_sz}+\text{H.c.},
\end{equation}
where $a_{mn}$ is the photon annihilation
operator for the $\text{TEM}_{mn}$ mode, $k_s=\omega_s /c=2\pi/\lambda_s$ is the wave vector of a plane wave approximating the $\text{TEM}_{00}$ mode propagating in the $z$ direction in a cavity section containing the atomic system, $\varepsilon_0$ is the permittivity of free space, and we suppose that the field is linear polarized along, e.g., the $x$ axis.
The corresponding input and output fields in the vicinity of the partially transmitting mirror are described in a similar way in an appropriate coordinate system. In this case, the operators $a_{mn}(t)$ should be replaced by $a_{mn}^{\text{in,out}}(t)$ differing by a factor of $\sqrt{L/c}$. The propagation of these fields outside the cavity may also be considered in the paraxial approximation, which is beyond the scope of our present work.
\begin{figure}
\includegraphics[width=8cm]{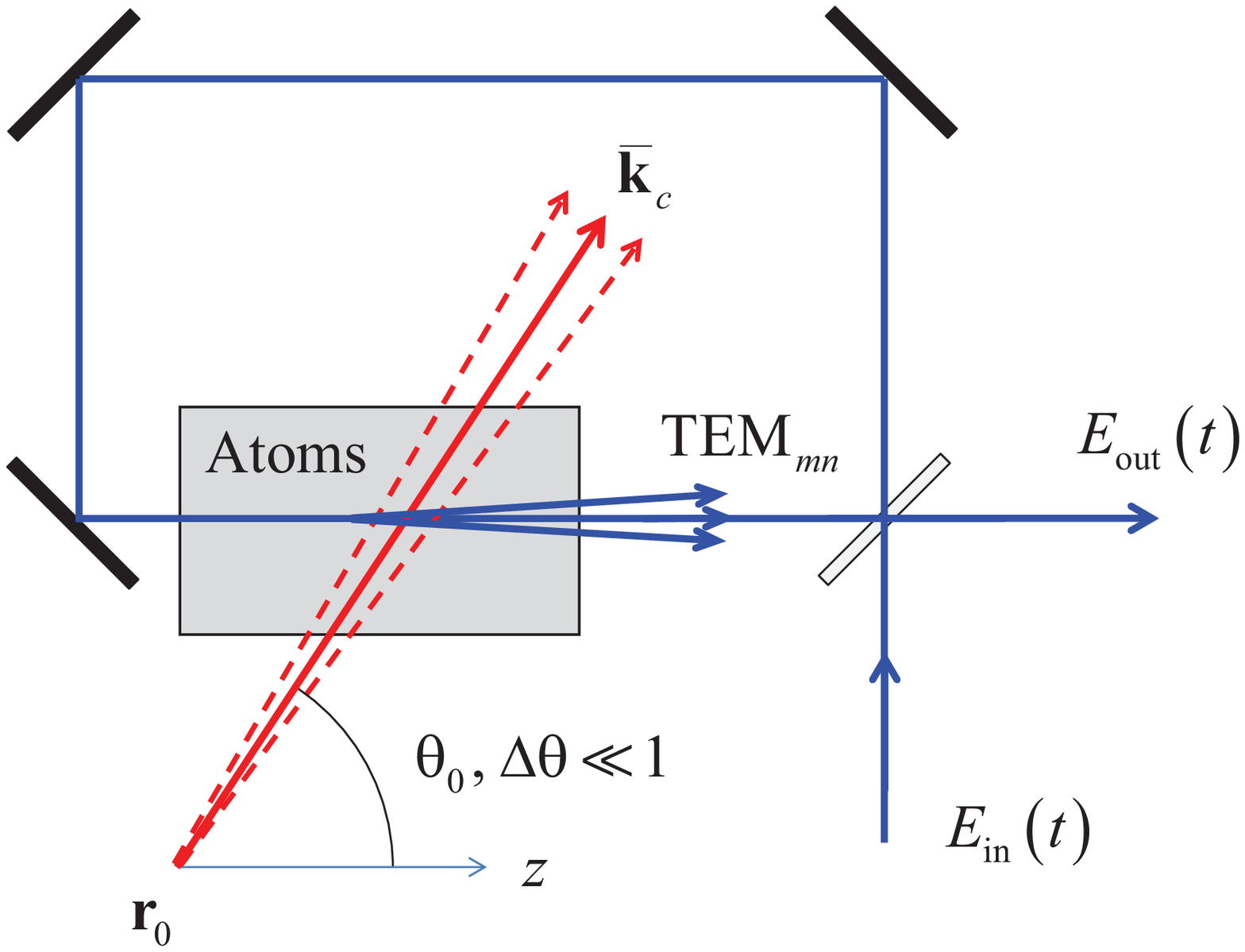}\\
\includegraphics[width=6cm]{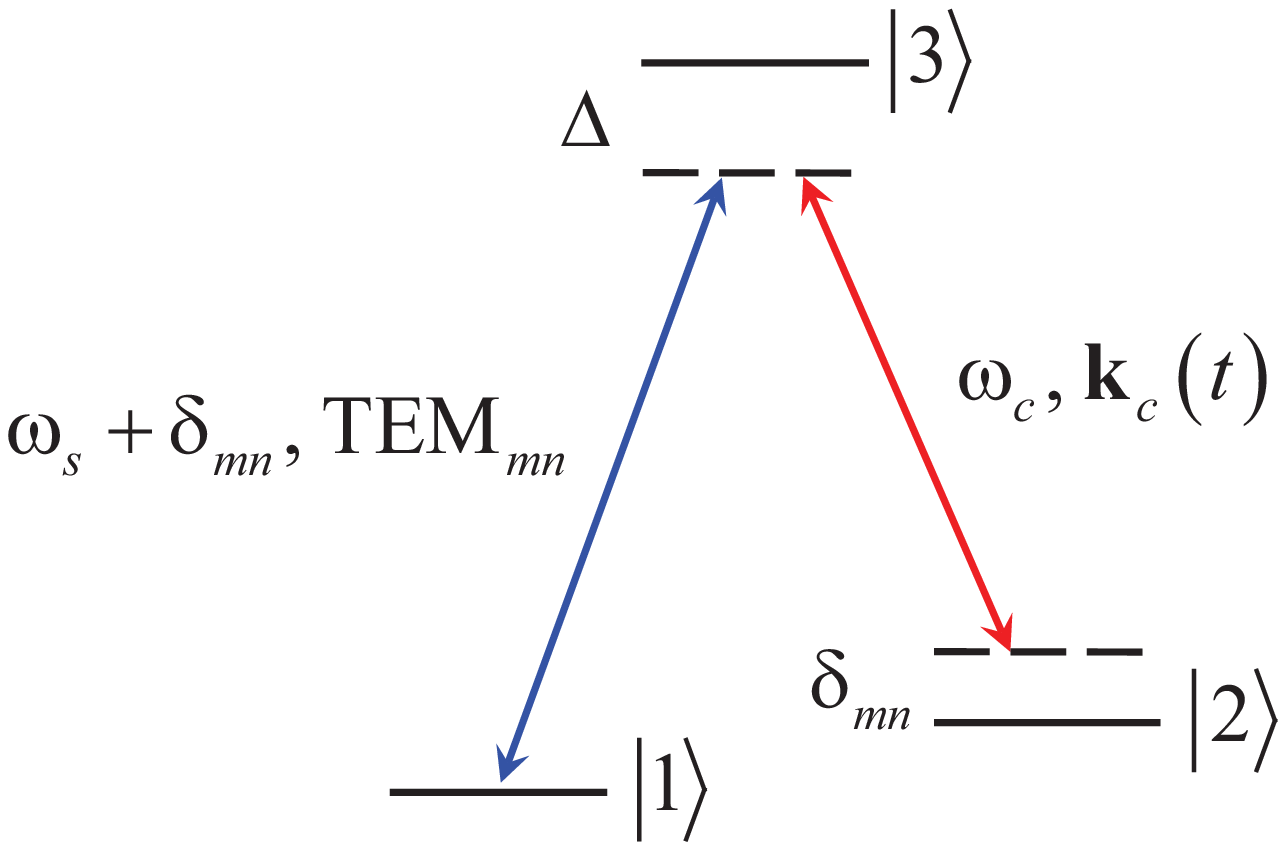}
\caption{\label{fig:levels} (Color online) Geometry (above) and energy diagram (below) illustrating cavity-assisted off-resonant Raman interaction between a strong control field with a wave vector $\vec{k}_c(t)$, a multimode signal field as a superposition the $\text{TEM}_{mn}$ cavity modes propagating along the $z$ axis, and a three-level atomic medium. $\vec{r}_0$ stands for the point where the phase shift of the control field induced by the rotation remains zero, $\bar{\vec{k}}_c$ corresponds to the average wave vector, and $\theta_0$ is the angle between $\bar{\vec{k}}_c$ and the $z$ axis. Mirrors are supposed to be forming a single-ended ring cavity for the signal field and fully transmitting for the control field.}
\end{figure}

The control field is supposed to be a monochromatic plane wave with a rotated wave vector (within a small angle), and its electric field is described as in \cite{ZKK_2013} by
\begin{equation}
E_c(\vec{r},t)=E_0\,e^{i[\bar{\vec{k}}_c\cdot\vec{r}-\omega_ct+\phi(\vec{r},t)]}+\text{c.c.},
\end{equation}
where $E_0$ is a constant amplitude of the plane wave, $\bar{\vec{k}}_c$ is an average value of the wave vector $\vec{k}_c(t)$ during the rotation ($k_c=\omega_c/c=2\pi/\lambda_c$), and $\phi(\vec{r},t)$ is a phase shift due to the rotation. The latter can be written as
\begin{equation}
\phi(\vec{r},t)=\vec{q}(t)\cdot(\vec{r}-\vec{r}_0),
\end{equation}
where
\begin{equation}
\vec{q}(t)=\int_{t_0}^t \frac{d\vec{k}_c(t)}{dt}\,dt
\end{equation}
is the net change of the wave vector from the moment of time $t=t_0$, when $\vec{k}_c(t)=\bar{\vec{k}}_c$, to the moment $t$. In what follows, we assume that $t_0=0$. The point $\vec{r}_0$ is referred to as a phase stationary point since the phase shift remains constant there. The product $\vec{q}(t)\cdot\vec{r}_0$  becomes equal to zero in two cases: if $\vec{r}_0=0$, i.e., when the phase stationary point is located at the center of the sample, and if $\bar{\vec{k}}_c$ is directed from the phase stationary point to the center of the sample [$\vec{k}_c(t)$ is rotated within a small angle so that in the first order $\vec{q}(t)\perp\bar{\vec{k}}_c$]. Under such conditions, we obtain
\begin{equation}\label{B2}
\phi(\vec{r},t)=\vec{q}(t)\cdot\vec{r}.
\end{equation}
The control field is also assumed to be linearly polarized, e.g., along the $y$ axis, and therefore propagated in the $(x,z)$ plane.

The atomic ensemble is supposed to fill the cavity in the transverse directions and to have a length $L_z\leq L$ along the cavity axis. Thus the sample can be approximated by a parallelepiped with cross section $A$ and volume $V_{a}=AL_z$. It should be noted that a standing-wave cavity can also be described in the present model by adding counterpropagating modes of the signal field. Then the case $L_z=L$ simply means a cavity filled by the atoms. The atoms are assumed to be motionless and prepared initially in the state $\ket{1}$. In addition, we assume that the time of propagation of photons through the system, which may be defined as $\sqrt[3]{V}/c$, is negligibly short compared to the evolution time of the slowly time-varying field amplitudes. In particular, $\phi(\vec{r},t)$ is considered as a slowly varying function on the propagation time scale.

The Hamiltonian of the three-level
system in the dipole and rotating wave approximations is
\begin{equation}
H=H_0+H_\text{int},
\end{equation}
where
\begin{align}
H_0=&\hbar\sum_{mn}(\omega_s+\delta_{mn})\, a_{mn}^\dag a_{mn}\\ &+\sum_{j=1}^{N}\left(
\hbar\omega_2\,\sigma_{22}^{j}+\hbar\omega_3\,\sigma_{33}^{j}\right),
\end{align}
\begin{align}
H_\text{int}=&-\hbar\sum_{j=1}^{N}\Omega\,\sigma_{32}^{j}\,e^{i\bar{\vec{k}}_c\cdot\vec{r}_j-i\omega_c
t+i\phi(\vec{r}_j,t)}\\
&-\hbar g\sum_{j=1}^{N}\sum_{mn}u_{mn}(\vec{r}_j)\,e^{ik_sz_j}\,a_{mn}\,\sigma_{31}^{j}+\text{H.c.}
\end{align}
Here $\sigma_{mn}^j=\ket{m_j}\bra{n_j}$ are the atomic operators,
$\ket{n_j}$ is the $n$th state ($n=1,2,3$) of the $j$th atom with the
energy $\hbar\omega_n$ ($\omega_1=0<\omega_2<\omega_3$), $\vec{r}_j$ is
the position of the $j$th atom, $\delta_{mn}$ is a frequency shift of the $\text{TEM}_{mn}$ mode relative to the basic frequency $\omega_s$, $\Omega=d_{23}E_0/\hbar$ is the Rabi
frequency of the classical field, $g=d_{13}\sqrt{\omega_s/2\varepsilon_0\hbar V}$ is the coupling constant
between the atoms and the signal field, and $d_{mn}$ is the dipole moment of the transition between the states $\ket{m}$ and $\ket{n}$, which is supposed to be real.

We follow the Heisenberg-Langevin approach,
which is typically used for studying Raman memories and particularly their cavity models (see, e.g., \cite{GALS_2007_1}, and references therein).
In the Heisenberg picture, we define the following slowly varying
atomic operators: $P_j=\sigma_{13}^{j}{\,e}^{i\omega_s t-ik_sz_j}$,
$S_j=\sigma_{12}^{j}{\,e}^{i(\omega_s-\omega_c)t-i(\vec{k}_s-\bar{\vec{k}}_c)\cdot\vec{r}_j}$, and cavity mode
amplitudes $\mathcal{E}_{mn}=a_{mn}\,e^{i(\omega_s+\delta_{mn}) t}$. The corresponding input
(output) field amplitude matched to the $mn$-th transverse mode is defined in a similar way and denoted as
$\mathcal{E}_{mn}^\text{in}$ ($\mathcal{E}_{mn}^\text{out}$). The input-output
relations (boundary conditions) for the single-ended cavity read \cite{CG_1984,GC_1985}
\begin{equation}\label{in_out}
\mathcal{E}_{mn}^\text{out}(t)=\sqrt{2\kappa_{mn}}\,\mathcal{E}_{mn}(t)-\mathcal{E}_{mn}^\text{in}(t),
\end{equation}
where $2\kappa_{mn}$ is the cavity decay rate. Assuming that all the population
is initially in the ground state and taking into account that the
signal field is weak, we obtain the following Heisenberg-Langevin equations:
\begin{align}
\dot{P_j}&=-(\gamma_P+i\Delta)P_j+i\Omega
S_j\,e^{i\phi(\vec{r}_j,t)}\nonumber\\
&\quad +ig\sum_{m,n}\mathcal{E}_{mn}\,e^{-i\delta_{mn}t}\,u_{mn}(\vec{r}_j),\label{1}\\
\dot{S}_j&=-(\gamma_S+i\Delta_S)S_j+i\Omega^\ast
P_j\,e^{-i\phi(\vec{r}_j,t)},\label{2}\\
\dot{\mathcal{E}}_{mn}&=-\kappa_{mn}\mathcal{E}_{mn}+\sqrt{2\kappa_{mn}}\,\mathcal{E}_{mn}^\text{in}(t)\nonumber\\
&\quad +ig\,e^{i\delta_{mn}t}\sum_j
P_j\,u_{mn}^\ast(\vec{r}_j).\label{3}
\end{align}
Here $\gamma_P$ and $\gamma_S$ are the rates of dephasing, which
in a general case include both homogeneous and inhomogeneous
broadening of the resonant transitions, $\Delta=\omega_3-\omega_s$
is a one-photon detuning, and $\Delta_S=\omega_2+\omega_c-\omega_s$
is a two-photon detuning. The Langevin noise
atomic operators are not included since they make no contribution to normally
ordered expectation values when almost all atoms remain in the ground state.

In the Raman limit, when the single-photon detuning is sufficiently large,
adiabatically eliminating $P_j$ in Eqs. (\ref{1})--(\ref{3}),
and going to the collective atomic operators
\begin{align}
S_{mnp}&=\frac{1}{\sqrt{N}}\sum_{j} S_{j}\,u_{mnp}^\ast(\vec{r}_j)\\
&=\frac{1}{\sqrt{N}}\int d\vec{r} S(\vec{r})\,\,n(\vec{r})\,u_{mnp}^\ast(\vec{r}),
\end{align}
where $n(\vec{r})$ is the atomic number density, we obtain
\begin{align}
\dot{S}_{m'n'p}&=-\gamma_R S_{m'n'p}\nonumber\\
&\quad+ig^\ast_R\sqrt{N}\sum_{mn}\mathcal{E}_{mn}\,e^{-i\delta_{mn}t}B^\ast_{mn,m'n'p}(t),\label{A1}\\
\dot{\mathcal{E}}_{mn}&=-\kappa_{mn}\mathcal{E}_{mn}+\sqrt{2\kappa_{mn}}\,\mathcal{E}_{mn}^\text{in}(t)\nonumber\\
&\quad+ig_R\sqrt{N}\,e^{i\delta_{mn}t}\sum_{m'n'p}B_{mn,m'n'p}(t)S_{m'n'p}.\label{A2}
\end{align}
Here
\begin{equation}\label{B}
B_{mn,m'n'p}(t)=\frac{1}{N}\int d\vec{r}\,n(\vec{r})\,e^{i\phi(\vec{r},t)}\,u^\ast_{mn}(\vec{r})\,u_{m'n'p}(\vec{r}),
\end{equation}
$\gamma_R=\gamma_S+\gamma_P|\Omega|^2/\Delta^2$,
$g_R=g\Omega/\Delta$, and the resulting frequency shift
$\Delta'_S=\Delta_S-|\Omega|^2/\Delta$ has been compensated by
tuning the coupling field frequency. In what follows, we take the atomic number density to be constant in space so that
\begin{equation}
n(\vec{r})=\left\{
             \begin{array}{ll}
               N/V_a, & \text{$\vec{r}\in V_a$} \\
               0 & \text{otherwise.}
             \end{array}
           \right.
\end{equation}

The standard figures of merit that describe quantum storage are total efficiency $\eta$ and fidelity $F$. The first is defined as
\begin{equation}
\eta=\frac{N_\text{out}}{N_\text{in}},
\end{equation}
where
\begin{align}
N_\text{in}&=\sum_{mn}\int_{-\infty}^0 dt\, \avr{\mathcal{E}_{mn}^{\text{in}\dag}(t)\mathcal{E}_{mn}^\text{in}(t)},\\
N_\text{out}&=\sum_{mn}\int_{0}^\infty dt\, \avr{\mathcal{E}_{mn}^{\text{out}\dag}(t)\mathcal{E}_{mn}^\text{out}(t)},
\end{align}
considering that the storage process terminates at the moment $t=0$, while the retrieval process begins at this moment of time.
The fidelity may be defined as
\begin{equation}
F=\eta F',
\end{equation}
where
\begin{equation}
F'=\frac{\left|\sum_{mn}\int_0^\infty dt\,\avr{\mathcal{E}_{mn}^{\text{out}\dag}(t)\mathcal{E}_{mn}^{\text{in}}(t-\bar{t})}\right|^2}{N_\text{in}N_\text{out}}
\end{equation}
is the correlation between the input and output pulse envelopes, and $\bar{t}$ is the delay that maximizes $F'$.

The presence of time-dependent phase $\phi(\vec{r},t)$ in the coupling coefficients (\ref{B}) leads, in a general case, to the cross-talk between different transverse modes of the field and atomic coherence, thereby reducing efficiency and fidelity of the multimode storage. In Sec.~III we discuss the conditions for making this cross talk negligible and parallel storage of a number of transverse modes possible.

\section{Storage and retrieval of multimode single-photon states}

Let the control field propagate at some angle $\theta_0$ to the signal field, i.e., $\bar{\vec{k}}_c=(k_c\sin\theta_0,0,k_c\cos\theta_0)$, and its wave vector is rotated within a small angle $\Delta\theta$ during the interval $T$ with a constant angular rate, so that $\vec{q}(t)=(k_c\cos\theta_0,0,-k_c\sin\theta_0)\frac{\Delta\theta}{T} t$. To be more specific, we consider Hermite-Gaussian modes and assume, for simplicity, that the origin of the $z$ axis coincides with the beam waist. Then
\begin{align}
u_{mn}(\vec{r})=&\frac{a\sqrt{A}}{\sqrt{\pi\,n!\,m!\,2^{n+m}}}\,H_m(ax)\,H_n(ay)\,e^{-a^2(x^2+y^2)/2}\nonumber\\
&\times e^{-ik_s(x^2+y^2)/2R(z)}\nonumber\\
&\times e^{i(m+n+1)\arctan(z/z_R)},
\end{align}
where $a=\sqrt{2}/w(z)$, $w(z)=w_0\sqrt{1+(z/z_R)^2}$ is the spot size (beam radius), $R(z)=z[1+(z_R/z)^2]$, $z_R=\pi w_0^2/\lambda_s$, and $w_0$ is the waist size. In addition, we suppose that the sample length $L_z$ is smaller than the confocal parameter $2z_R$ so that $w(z)\approx w_0$ and $\arctan(z/z_R)\approx z/z_R$ within the sample.
Under such conditions,
\begin{align}\label{B3}
B_{mn,m'n'p}(t)&=\text{sinc}[(t+t_p+t_{mn,m'n'}){\pi}/{\delta}]\nonumber\\&\times\frac{1}{A}\iint dx\,dy\,e^{iq_x(t)x} u^\ast_{mn}(x,y)\,u_{m'n'}(x,y),
\end{align}
where
\begin{equation}
\delta=\frac{T}{\Delta\theta}\frac{\lambda_c}{L_z\sin\theta_0},
\end{equation}
$t_p=p\,\delta L_z/L$, and $t_{mn,m'n'}=[(m'-m)+(n'-n)]\,\delta L_z/(2\pi z_R)$. The time interval $\delta$ corresponds to switching between two orthogonal longitudinal spin modes (with $z$ components of wave vectors $2\pi p/L_z$, $p\in \mathbb{Z}$ relative to $\vec{k}_s-\bar{\vec{k}}_c$), which happens once the angle of rotation exceeds the effective diffraction angle $\lambda_c/(L_z\sin\theta_0)$. Since $L_z\leq L$, the interval between wave vectors of the spin modes $2\pi/L_z$ may be larger than that of the cavity modes $2\pi/L$. Therefore, switching between the spin modes needs the index $p$ to be changed on the value $L/L_z$, which is greater than 1. The spatial multimode storage is possible provided that switching between the longitudinal spin modes occurs much faster than the cross talk between transverse modes, i.e., when $q_x(t)x$ remains small within the rotation time $T$. For small values of $q_x(t)x$, when $e^{-iq_x(t)x}\approx 1-iq_x(t)x$, we have
\begin{align}\label{nnmm}
\frac{1}{A}\iint &dx\,dy\,e^{iq_x(t)x} u^\ast_{mn}(x,y)\,u_{m'n'}(x,y)\\&\approx
(\delta_{mm'}-i\alpha_{\pm}(t)\,\delta_{m\pm 1,m'})\,\delta_{nn'},
\end{align}
where $\alpha_+(t)=\frac{q_x(t)w_0}{2}\sqrt{m+1}$ and $\alpha_-(t)=\frac{q_x(t)w_0}{2}\sqrt{m}$. Since $q_x(T)=2\pi(T/\delta)(\cot\theta_0/L_z)$, we obtain $|\alpha_{\pm}(T)|\ll 1$ provided that
\begin{equation}\label{maincondition}
\left|\tan\theta_0\right|\gg\frac{\pi w_0}{L_z}\frac{T}{\delta}\sqrt{m+1}.
\end{equation}
Otherwise, integration in Eq.~(\ref{nnmm}) yields an oscillating function that decays in time. The decay time decreases with increasing $\cos\theta_0$, and once it approaches $T$, the storage efficiency and fidelity are reduced. According to Eq.~(\ref{maincondition}), small angles between the control field and signal field beams are only possible for pencil-like geometries, when $L_z\gg w_0$, and for low order modes.

Under the condition (\ref{maincondition}), the coupling coefficient (\ref{B3}) is approximated by
\begin{equation}
B_{mn,m'n'p}(t)=\text{sinc}[(t+t_p){\pi}/{\delta}]\,\delta_{mm'}\,\delta_{nn'},
\end{equation}
and Eqs.~(\ref{A1}) and (\ref{A2}) take the form
\begin{align}
\dot{S}_{mnp}&=-\gamma_R S_{mnp}\nonumber\\
&\quad+ig^\ast_R\sqrt{N}\,\mathcal{E}_{mn}\,e^{-i\delta_{mn}t}\,\text{sinc}[(t+t_p)\pi/\delta],\label{A1a}\\
\dot{\mathcal{E}}_{mn}&=-\kappa_{mn}\mathcal{E}_{mn}+\sqrt{2\kappa_{mn}}\,\mathcal{E}_{mn}^\text{in}(t)\nonumber\\
&\quad+ig_R\sqrt{N}\,e^{i\delta_{mn}t}\sum_{p}\,\text{sinc}[(t+t_p)\pi/\delta]S_{mnp}.\label{A2a}
\end{align}
Thus at different moments of time the signal field effectively interacts with spin waves having different longitudinal components of wave vectors. The switching of the collective atomic-field interaction from one spin wave to another by a rotated control field is possible in an extended resonant medium due to a phase matching condition. As a result, absorption and emission of the field is accomplished through reversible mapping of its amplitude into a superposition of the spin waves forming coherence spatial grating. On the other hand, as was shown in \cite{ZKK_2013}, the off-resonant Raman interaction with a rotated transverse control field is mathematically equivalent to GEM. The latter provides the physics of the proposed scheme in the frequency domain.

According to Eqs.~(\ref{A1a}) and (\ref{A2a}), different transverse modes evolve independently of each other, and following \cite{KK_2011}, we can solve these equations analytically. Let the atomic system interact with
the quantum field during the time interval $[-T,0]$ with the
initial condition $S_{mnp}(-T)=0, \forall m,n,p$. We are interested in parallel storage and retrieval of single-photon wave packets that correspond to different transverse modes and have duration $\delta_p$ smaller than $T$, but larger than $\delta$. Assuming that the cavity field $\mathcal{E}_{mn}$ varies slowly during
$\delta$, and $\gamma_R,\delta_{mn}\ll \delta^{-1}$, by the end of the storage process from Eqs.~(\ref{A1a}) and (\ref{A2a}) we derive
\begin{align}
S_{mnp}(0)&=\frac{ig^\ast_R\sqrt{N}\sqrt{2\kappa_{mn}}\,\delta}{\kappa_{mn}+\Gamma}\,\mathcal{E}_{mn}^\text{in}(t_p)\,e^{(-i\delta_{mn}+\gamma_R)
t_p},\label{C}
\end{align}
where
\begin{equation}\label{Gamma}
2\Gamma=|g_R|^2N\delta
\end{equation}
is the rate of the cavity-assisted collective atomic transition.
Equation (\ref{C}) describes the mapping of an input single-photon
wave packet corresponding to the $\text{TEM}_{mn}$ mode into a superposition of collective excitations (spin
waves) with different longitudinal components of wave vectors. It is valid provided that $\Gamma+\kappa_{mn}$ is much
greater than the bandwidth of the input field (see \cite{KK_2011} for details), which corresponds to the bad-cavity limit.
To what extent the value $\Gamma+\kappa_{mn}$ may be reduced with respect to the pulse bandwidth is discussed below in this section.

Retrieval is achieved by off-resonant interaction of the
atomic system with the control field when the values of $\vec{k}_c(t)$ that
are used for storage are scanned again. Let the control field be applied during the time interval
$[0,T]$, when $\mathcal{E}_{mn}^\text{in}(t)=0$, and the direction of its rotation is reversed. In this
case, by solving Eq.~(\ref{A2a}) with the initial condition (\ref{C}), and using Eq.~(\ref{in_out}), we obtain
\begin{equation}\label{D}
\mathcal{E}_{mn}^\text{out}(t)=-\frac{2\Gamma}{\kappa_{mn}+\Gamma}\mathcal{E}_{mn}^\text{in}(-t)\,e^{-2\gamma_Rt}.
\end{equation}
The output field becomes a time-reversed
replica of the input field provided that the duration of the signal pulse is much smaller than the decay time $1/\gamma_R$, and  the
efficiency of the storage followed by retrieval is maximum under
impedance-matching condition $\kappa_{mn}=\Gamma$ \cite{AS_2010,MAG_2010}. Thus the less the dispersion of the cavity decay rates $\kappa_{mn}$, the less the distortion of the retrieved spatial multimode state. In fact, we need diffraction losses to be smaller than those through the partially transmitting mirror.
Another important condition for effective multimode storage and retrieval, which was used in obtaining (\ref{D}), is $\delta_{mn}\delta\ll 1$. The time interval $\delta$ in the present scheme plays the same role as reversal inhomogeneous linewidth in photon-echo-based schemes, and in order to effectively store and reconstruct the input pulse, we need the absorption bandwidth to be larger than the input spectrum. Since the latter is determined by both pulse shape and its frequency detuning, not only the amplitude $\mathcal{E}_{mn}$, but also the phase factor $e^{i\delta_{mn}t}$ should be smooth with respect to $\delta$. These features are illustrated in Fig.~2. According to the numerical solutions of Eqs.~(\ref{A1a}) and (\ref{A2a}), a Gaussian pulse with a duration $\delta_p$ (full width at half maximum) as short as $\delta$ or frequency detuning as large as $0.6/\delta$ can be stored and recalled with the efficiency 0.99. It means that spatial multimode states corresponding to some spectral range of transverse cavity modes can effectively be stored if their duration is shorter than the reversal spectral range of the involved modes. Taking $T=5\delta$, we find that the minimum angle of the control-field rotation needed for storage of a single pulse with near 100\% efficiency is five times larger than the diffraction angle $\lambda_c/(L_z\sin\theta_0)$. This value is the minimum resolvable spot number which should be provided by a beam deflector during the angular scanning.
\begin{figure}
\includegraphics[width=8.6cm]{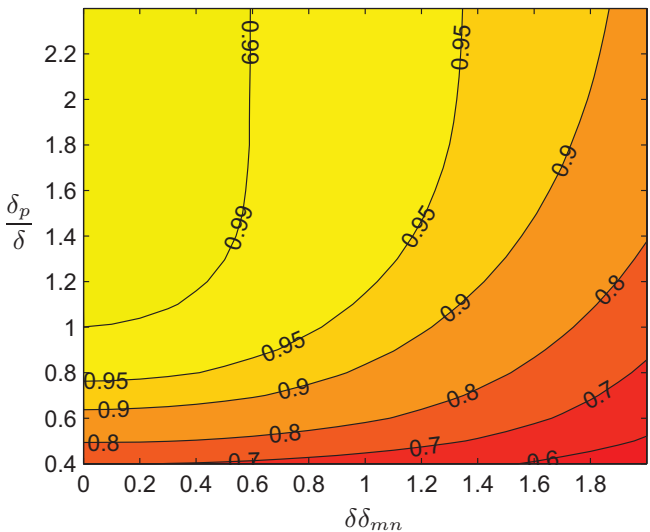}
%\put(-125,0){$\delta\delta_{mn}$} \put(-240,100){$\dfrac{\delta_p}{\delta}$}
\caption{\label{fig:Efficiency} (Color online) Total efficiency of storage followed by retrieval $\eta$
for a Gaussian pulse of duration $\delta_p$ in a transverse mode with a frequency detuning $\delta_{mn}$
for different values of $\delta$. The plot is
obtained by numerically solving Eqs.~(\ref{A1a}) and (\ref{A2a}),
treated as complex number equations, with the conditions
$\theta_0=\pi/2$, $L=L_z$, $\kappa\delta=4.2$, $T=30\delta$ and $\gamma_R=0$.}
\end{figure}

Figure~3 illustrates the total efficiency of quantum storage as a function of the cavity decay rate $\kappa_{mn}$ for different values of the pulse duration $\delta_p$. The maximum efficiency is achieved at some optimal ratio between the switching time $\delta$ and the cavity decay rate $\kappa_{mn}$, namely when $\kappa_{mn}\delta\approx 5$ for $\delta_p\geq 2\delta$. The efficiency gradually (rapidly) decreases when $\kappa_{mn}$ becomes higher (smaller) than the optimal value, and becomes less sensitive to $\kappa_{mn}$ with increasing pulse duration. Since the reversal switching time $\delta$ is equivalent to the inhomogeneous linewidth of the atomic transition, this optimal ratio is equivalent to the spectral matching condition discussed in \cite{MAG_2010}.
\begin{figure}
\includegraphics[width=8.6cm]{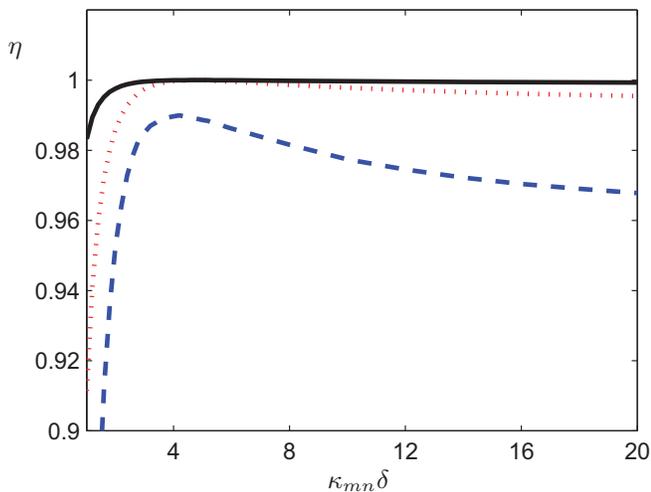}
%\put(-130,0){$\kappa_{mn}\delta$} \put(-240,150){$\eta$}
\caption{\label{fig:Efficiency2} (Color online) Total efficiency of storage followed by retrieval $\eta$
of a Gaussian pulse of duration $\delta_p=\delta$ (blue dashed line), $\delta_p=2\delta$ (red dotted line), and
$\delta_p=5\delta$ (black solid line) for different values of the cavity decay rate $\kappa_{mn}$. The plots are
obtained by numerically solving Eqs.~(\ref{A1a}) and (\ref{A2a}),
treated as complex number equations, with the conditions
$\theta_0=\pi/2$, $L=L_z$, $\delta_{mn}=0$, $T=30\delta$ and $\gamma_R=0$.}
\end{figure}

It should be noted that, as in the case of refractive index control \cite{KK_2011}, a single-photon wave packet may be reconstructed without time
reversal if $\vec{k}_c(t)$ is rotated
during retrieval in the same direction as during storage. In this case we obtain
\begin{equation}\label{D1}
\mathcal{E}_{nm}^\text{out}(t)=-\frac{2\Gamma}{\kappa_{mn}+\Gamma}\mathcal{E}_{mn}^\text{in}(t-T)\,e^{-\gamma_RT},
\end{equation}
which means that temporal shape of the pulse is not deformed by the
dephasing process.

Finally, let us estimate the number of transverse modes which can be stored and recalled simultaneously and the number of pulses which can be stored in a sequence. As was mentioned above, for effective multimode storage the diffraction losses should be much smaller than those through the partially transmitting mirror. In addition, small frequency intervals $\delta_{mn}$ between the transverse modes are preferable. To be more precise, according to the results of numerical simulations presented above, high storage efficiency is achieved provided that $\delta_{mn}\ll\delta^{-1}\ll 2\kappa_{mn}$, which means that the cavity modes may be nondegenerate but should be nonresolvable in the frequency domain. From this point of view, confocal cavities seem to be the optimal ones. In this case, the diffraction losses per round trip can be estimated by \cite{I_1980}
\begin{equation}
\alpha_{mn}=1-(1-\alpha_m)(1-\alpha_n),
\end{equation}
where
\begin{equation}
\alpha_m=4\sqrt{\pi}\,\frac{1}{m!}\,8^m (2\pi N_F)^{m+1/2}\,e^{-4\pi N_F},
\end{equation}
and $N_F=A/(\lambda L)$ is the cavity Fresnel number. If we take the mirror transmittance of 0.1\%, thereby requiring $\alpha_{mn}\lesssim 10^{-4}$, and $N_F=10$, then the maximum value of the transverse indexes proves to be about 30, and the total number of accessible transverse modes approaches $10^3$.
Storage of such a multimode transverse field can be combined with that of the multimode longitudinal profile, i.e., with storage of a complicated pulse shape or a sequence of pulses. From Eq.~(\ref{maincondition}), taking $\pi w_0/L_z=0.02$, $T/\delta=5$, and requiring the left-hand side to be an order larger than the right-hand side, for $m=30$ we obtain $80^\circ\leq\theta_0\leq 100^\circ$. Such an angular scanning range allows one to store about 100 of the pulses in a sequence, depending on the minimum value of rotation angle per pulse $\sim 5\lambda_c/L_z$. Thus we can predict a large storage capacity for the proposed memory scheme.

\section{Implementation of the scheme in a solid-state material}

Quantum storage based on off-resonant Raman interaction has been successfully demonstrated in warm atomic vapors of $\text{Rb}$ atoms \cite{HHSOLB_2008,HSHLLB_2009,HSCLB_2011,HSRPHLB_2012} and $\text{Cs}$ atoms \cite{RNLSLLJW_2010,RMLNLW_2011,RNJMCELKLW_2012}. In such media, the broadening of the two-photon spin transition due to the thermal motion of atoms can be minimized by working in collinear geometry (when control and signal fields are copropagating) and by restriction the motion of atoms using a buffer gas. Increasing the angle between the control and signal fields leads to significant residual Doppler broadening. In addition, a small spatial period of coherence grating in the case of non-collinear excitation makes the storage process more sensitive to the diffusion of the atoms. Therefore, an ensemble of cold atoms trapped in an optical lattice or impurity atoms in a solid-state material seem to be the most promising storage media for the present memory scheme.
In this section, we discuss the possibility of implementation of the proposed scheme in solids. One promising candidate---an ensemble of defect centers in diamond such as nitrogen-vacancy centers---was discussed already in \cite{ZKK_2013}. Here we focus on materials with much weaker optical transitions, namely, crystals doped by rare-earth ions, thereby making use of the advantages of cavity-assisted interactions.

As an example, we consider low-strain crystals $\text{YLiF}_4$ or $\text{LuLiF}_4$ doped by $\text{Er}^{3+}$ ions. In particular, an isotopically pure crystal of $\text{YLiF}_4$ (when only ${}^7\text{Li}$ is present) demonstrates an extremely narrow inhomogeneous linewidth (as low as 15~MHz at 0.005 at.\%) for optical transitions of $\text{Er}^{3+}$ ions \cite{MCM_1992,CPKA_2000,TBC_2011}. Moreover, $\text{Er}^{3+}$ ions present a hyperfine structure due to ${}^{167}\text{Er}^{3+}$ isotope, which allows one to identify $\Lambda$-type level structures for storage \cite{BBMBLCLBGGG_2010}. These features make such crystals very promising for implementation of the cavity-assisted Raman scheme. For the electronic transition ${}^4\text{I}_{15/2}(1)-{}^4\text{I}_{13/2}(1)$ ($\lambda=1530$~nm), we have oscillator strength $f=2\times 10^{-7}$ \cite{TBC_2011}. Then, taking concentration of impurities $7\times 10^{17}~\text{cm}^{-3}$ (0.005 at.\%) and assuming that the atomic system fills the cavity, we obtain $g^2N=5\times 10^{19}~\text{s}^{-2}$. The impedance-matching condition $\kappa=\Gamma$ may be achieved, e.g., with the following values of parameters: $\kappa= 10^8~\text{s}^{-1}$, $\delta=2\times 10^{-7}~\text{s}$, and $(\Omega/\Delta)^2=2\times 10^{-5}$. For $\Delta/2\pi = 100~\text{MHz}$ we need $\Omega/2\pi=4.5\times 10^5~\text{Hz}$, which corresponds to the intensity of the control field $\sim 65~\text{W/cm}^2$. Considering a standing-wave cavity, we can take $L_z=2.5~\text{mm}$, and a control-field beam of 2.5~mm diameter. Then we need the transmittance of the output cavity mirror of 0.5\%, and the control-field power about 1.6~W. Such a regime can be realized with commercial cw fiber lasers.
%To use pulsed lasers we can take $\delta=2\cdot 10^{-9}~\text{s}$ and samples with 0.1\% concentration of impurity ions, where linewidth of 100~MHz is %expected in the isotopically pure crystals. In this case, taking $\Delta/2\pi=1~\text{GHz}$, we need the control field intensity $\sim %33~\text{kW/cm}^2$.
Regarding the control-field angular scanning, typical resolvable spot numbers of 13 \cite{SSGGCSRMCK_2002,HSK_2005,BBBB_2009} are achievable at the deflection period of 61.5~ps \cite{HSK_2005} with electro-optic laser beam deflectors.

\section{Conclusion}
It is shown that spatial multimode single-photon wave packets can be stored and
recalled in a resonant three-level medium placed in a cavity by means of continuous phase-matching control.
The proposed scheme generalizes those developed previously in \cite{KK_2011,ZKK_2013}. In order to reversibly map a spatial multimode state into a Raman coherence grating, it is necessary to rotate the wave vector of the control field so that wave vectors of the spin coherence can be of different longitudinal components with respect to the paraxial signal field. We show that the suggested cavity-assisted scheme may be implemented not only in gases, which are currently used for the Raman memory, but also in solid-state materials. The control-field angular scanning, which is used for controlling phase-matching, allows one to store and recall a sequence of weak optical pulses with a multimode transverse field, thereby providing a large storage capacity.

\section{Acknowledgments}

The work was supported by the NSF (Grant No.~0855688), RFBR (Grant
No.~12-02-00651-a), and the Program of the Presidium of RAS "Quantum mesoscopic and disordered structures".

%\bibliography{Thesis}

\end{document}